\documentclass[conference,letterpaper]{IEEEtran}

\usepackage[utf8]{inputenc} 
\usepackage[T1]{fontenc}
\usepackage{url}
\usepackage{ifthen}
\usepackage{cite}
\usepackage[cmex10]{amsmath} 
                             
\usepackage[table]{xcolor}
\usepackage{amssymb,amsthm}
\usepackage{subcaption}
\usepackage{tikz}
\usepackage{hyperref}
\usepackage{tabularx,environ}
\usetikzlibrary{backgrounds,calc,shapes,arrows}

\begin{document}
	
\title{Probabilistic Shaping in MIMO: Going Beyond 1.53dB AWGN Gain With the Non-Linear Demapper}

 \author{%
	\IEEEauthorblockN{Kirill Ivanov, Wei Yang, Jing Jiang}
	\IEEEauthorblockA{Qualcomm Technologies, Inc., San Diego, CA, 92121\\
		Email: \{kivanov, weiyang, jingj\}@qti.qualcomm.com}
}

\maketitle

\begin{abstract}
	Constellation shaping is a well-established method to improve upon a regular quadrature amplitude modulation (QAM). It is known that the gain achieved by any shaping method for an additive white Gaussian noise (AWGN) channel is upper-bounded by 1.53dB. However, the situation becomes less clear in the multiple-input and multiple-output (MIMO) setting.

	In this paper, we study the application of probabilistic shaping for MIMO channels. We utilize an efficient near-optimal demapper based on sphere decoding (SD) and demonstrate that it is possible to achieve more than 2dB gains, breaking the AWGN limit. It becomes possible because both signal and interference are shaped and the non-linear methods can capture this property and leverage on it to improve the demodulation performance.
	
\end{abstract}

\section{Introduction}
\IEEEPARstart{I}{n} modern wireless communication systems, a high data throughput is achieved by sending more than one coded bit per modulation symbol. A common practice is to utilize $2^{2M}$-QAM where the constellation points are arranged in a quadratic grid, which combines the improved spectral efficiency with the demodulation simplicity. However, its capacity falls short of the Shannon limit due to the regular structure.

Constellation shaping is a method to bridge the capacity gap by modifying the constellation to better match the capacity-achieving distribution of the channel. An upper bound on the gain achievable by any shaping method on AWGN channel is limited to 1.53 dB \cite{forney1989multidimensional}. In case when all points are equiprobable, the gain is attainable only with infinite-dimensional constellations, which severely limits its practical application. Instead, a more efficient method is to utilize probabilistic shaping (PS) with non-uniform signaling, i.e. to ensure that the constellation points follow a Maxwell-Boltzmann (MB) distribution \cite{kschischang1993optimal}. Various versions of PS have been proposed over the years, including schemes based on turbo \cite{raphaeli2004constellation} and LDPC \cite{kaimalettu2007constellation} codes. Alternatively, geometric shaping (GS) targets the same gains by making the constellation irregular. This method is utilized in ATSC 3.0 \cite{ATSC-A322} and DVB \cite{DVB-NGH2013} standards, albeit the lack of lattice structure in the constellation makes demodulation more complex.


A probabilistic shaping scheme proposed in \cite{bocherer2015bandwidth} is based on a distribution matcher (DM) that converts uniform data bits to modulation symbols with the target distribution together with a systematic error-correcting code and is shown to have good performance over a practical coded modulation system. PS (a.k.a., probabilistic amplitude shaping (PAS) in \cite{bocherer2015bandwidth}) is capable of achieving near-optimal shaping gains using constant composition distribution matcher (CCDM) \cite{schulte2016constant} and low-density parity-check (LDPC) codes at moderate to long block lengths even in case of the bit-interleaved coding modulation (BICM) \cite{caire1998bicm}, when the low-complexity receiver performs an independent demodulation of bits in the constellation symbol. Several more advanced distribution matchers were proposed to improve the performance over CCDM (especially in small DM block length region), such as energy-based shaping \cite{liu2023energy} and multiset-partition distribution matching \cite{fehenberger2019multiset}, at the cost of moderate complexity increase while enabling the operation at smaller block length to facilitate implementation.

As the technological progress goes forward, demands for even higher throughput can be met by utilizing several transmitting and receiving antennas. MIMO communication systems use spatial multiplexing of several data streams and benefit from the increased capacity and diversity. However, the complexity of maximum likelihood (ML) demodulation grows exponentially with the number of antennas which limits its practical applications and simple but suboptimal linear minimum mean square error (MMSE) method is often deployed instead. Nevertheless, the topic of developing practical near-optimal demappers has been extensively researched over the recent years. Notable examples include 
lattice reduction \cite{yao2002lattice,gan2009complex} and sphere decoder \cite{pohst1981computation,schnorr1991lattice,jalden2005complexity,studer2008soft}.

While multi-antenna techniques already found their way to the modern telecommunication standards, such as 5G \cite{3gpp.38.211} and Wi-Fi \cite{IEEE80211n} networks, probabilistic shaping is yet to be adopted and most academic works are focused on the case of AWGN channel. In case of Rayleigh fading, the combination of probabilistic shaping with BICM is no longer capacity-achieving and multi-level coding is needed \cite{iscan2018probabilistically}. Much less is known for MIMO channels. Some modest gains have been observed for 16-QAM \cite{kang2022probabilistic,bobrov2023probability} with MMSE detector.

In this paper, we demonstrate for the first time that shaping in MIMO channels can achieve gains substantially larger than 1.53dB  -- the maximal shaping gain obtainable over an AWGN channel. Moreover, we claim that having a non-linear detector such as sphere decoder is the crucial component. Our explanation for this phenomenon is the interference shaping: when the symbols on each spatial layer are shaped, it becomes easier for the detector to tell apart the superimposed symbols from different layers. One evidence in favor of our claim is the substantial reduction in the average number of nodes visited during the sphere decoder run when PS is enabled. Another comes from looking directly at the MIMO detector output and observing that the demapper performance for the layer improves when the other layer is shaped. However, if the interference from other layers is simply treated as noise, the receiver cannot leverage on its structure. Our results demonstrate that probabilistic shaping is a promising candidate for the next-generation networks such as 6G, bringing substantial performance improvements for the multi-antenna systems. 

\section{Background}

\subsection{Probabilistic amplitude shaping}
Consider $2^M$-PAM constellation 
$$
\mathcal O_R=\{\pm 1,\pm 3,\dots,\pm (2^M-1)\}.
$$ 
It is well-known that for AWGN channel the capacity-achieving input distribution is Gaussian. For finite constellations such as $\mathcal O_R$, capacity can be approached if the points are selected for transmission according to the Maxwell-Boltzmann distribution \cite{kschischang1993optimal}
\begin{equation}
\label{eq:MB}
P_{\nu}(x)=Ze^{-\nu ||x||^2}, x\in \mathcal O_R,
\end{equation}
where $Z$ is the normalization factor selected to ensure that $\sum_xP_{\nu}(x)=1$ and the parameter $\nu$ determines the source entropy.

Consider now an $(N, K)$ error-correcting code of rate $K/N=(M-1)/M$ with a systematic generator matrix
$$
\mathbf G=\begin{pmatrix}
	\mathbf I_k & \mathbf P
\end{pmatrix}.
$$
The encoding operation produces a length-$N$ codeword $\mathbf c=\begin{pmatrix}
	\mathbf d_K & \mathbf p_{N-K}
\end{pmatrix}$, where the first $K$ bits are equal to the data bits and hence have an identical distribution, whereas the remaining $N-K$ bits can be considered uniform \cite{bocherer2015bandwidth}. Since $P_{\nu}(x)=P_{\nu}(-x)$, we can map the data bits to the amplitudes of constellation symbols and the parity bits to the signs, i.e. $\mathbf d_K$ is decomposed into $K/(M-1)$ groups $A_i$ each containing $M-1$ bits, $\mathbf p_{N-K}$ is decomposed into $K/(M-1)$ singletons $P_i$ and we map the pair $(A_i,P_i)$ to the elements of $\mathcal O_R$ using Gray labeling. Note that the generalization to $2^{2M}$-QAM, defined as the direct product of two $2^{M}$-PAM constellations, is straightforward. In case when the code rate is greater than $(M-1)/M$, some information bits are left unshaped and mapped to the sign bits. The only remaining ingredient is how to generate the vector $\mathbf d_K$ so that the empiric distribution of amplitudes $A_i$ is close to \eqref{eq:MB}.

In this paper, we consider constant composition distribution matcher (CCDM) \cite{schulte2016constant}. It takes $K_c$ uniform data bits as an input and outputs $N_c$ amplitudes $A_i$ that follow the distribution $P_{\nu}$. For an amplitude sequence $A^{N_c}$, define $\mathcal T_x^{N_c}$ as the set of all permutations of $A^{N_c}$. The elements of $\mathcal T_x^{N_c}$ have the same empirical distribution $P(x)=\frac{|i: A_i=x|}{N_c}$, which leads to the following procedure\cite{bocherer2015bandwidth}:
\begin{enumerate}
	\item Compute $N_x=\left[N_c\cdot P_{\nu}(x)\right], x\in \{1,3,\dots,(2^M-1)\}$
	\item Select a sequence $A^{N_c}$ with an empirical distribution $P(x)=N_x/N_c$ and construct the set $\mathcal T_x^{N_c}$
	\item Compute the input length 
	$$
	K_c=\left\lfloor\log_2\frac{N_c!}{N_1!N_3!\dots N_{2^M-1}!}\right\rfloor
	$$
	\item Pick the set $\mathcal C_{CCDM}$ of $2^{K_c}$ sequences from $\mathcal T_x^{N_c}$ and define a bijective mapping $\{0,1\}^{K_c}\to \mathcal C_{CCDM}$.
\end{enumerate}
The mapping proposed in \cite{schulte2016constant} is based on the arithmetic coding and generalizes the idea from \cite{ramabadran1990coding} to the non-binary case. 

\subsection{MIMO}
In this and the next sections, we mostly follow the notations from \cite{studer2008soft}. Consider the complex-valued set $\mathcal O$ of constellation points s.t. 
\begin{equation}
	\label{eq:unitconst}
	\mathbb E_{s\in\mathcal O}[||s||^2]=1.
\end{equation}

A $M_t\times M_r$ MIMO system with $M_t$ transmit antennas and $M_r$ receive antennas can be characterized using the input-output relation 
\begin{equation}
\label{eq:mimo}
\mathbf y=\mathbf H\mathbf s + \mathbf n,
\end{equation}
where $\mathbf H\in \mathbb C^{M_t\times M_r}, \mathbf n\sim\mathcal{C}\mathcal N(0,\sigma \mathbf I_{M_r})$ and $\mathbf s\in \mathcal O^{M_t}$. In what follows, we assume that the set $\mathcal O$ corresponds to $2^{2M}$-QAM constellation \footnote{In case of PS the set $\mathcal O$ needs to be scaled appropriately so that \eqref{eq:unitconst} is satisfied for MB distribution.} and the receiver has perfect knowledge of the matrix $\mathbf H$. 

In a coded MIMO system, soft-output demapper computes the log-likelihood ratios (LLRs) $L_{j,b}$ for each bit $b$ in $j$-th detected symbol, which are subsequently passed to the channel decoder. Assuming the unit noise power, max-log-MAP LLRs can be computed as \cite{hochwald2003achieving}
\begin{equation}
\label{eq:mlm}
L_{j,b}(\mathbf y)=\min_{\mathbf s\in \mathcal{X}_{j,b}^{(0)}} ||\mathbf y-\mathbf H\mathbf s||^2-\min_{\mathbf s\in \mathcal{X}_{j,b}^{(1)}} ||\mathbf y-\mathbf H\mathbf s||^2,
\end{equation}
where $\mathcal{X}_{j,b}^{(i)}$ are the subsets of $\mathcal O^{M_t}$ that correspond to $b$-th bit of symbol $s_j$ being equal to $i$.

\subsection{Sphere decoder}
\label{ss:sd}
MIMO detection problem can be transformed into a tree search problem using the QR decomposition of the channel matrix $\mathbf H $ \cite{pohst1981computation,schnorr1991lattice,jalden2005complexity}. We obtain $\mathbf H=\mathbf Q \mathbf R$, where $M_r\times M_t$ matrix $\mathbf Q$ is unitary and $M_t\times M_t$ upper-triangular matrix $\mathbf R$ has real-valued positive entries on its diagonal. Multiplying both sides of \eqref{eq:mimo} by $\mathbf Q^H$ leads to the alternative formulation 
\begin{equation}
	\label{eq:mlm2}
	L_{j,b}(\mathbf{\tilde y})=\min_{\mathbf s\in \mathcal{X}_{j,b}^{(0)}} ||\mathbf{\tilde y}-\mathbf R\mathbf s||^2-\min_{\mathbf s\in \mathcal{X}_{j,b}^{(1)}} ||\mathbf{\tilde y}-\mathbf R\mathbf s||^2,
\end{equation}
where $\mathbf{\tilde y}=\mathbf Q^H\mathbf y$. Using the upper-triangular structure of matrix $\mathbf R$, we define partial distances
\begin{align}
	\label{eq:ped}
	d_i(\mathbf s)&=d_{i+1}(\mathbf s)+e_i(\mathbf s),i=M_t-1,\dots,0,\\
	e_i(\mathbf s)&=\left|\tilde y_i-\sum_{j=i}^{M_t-1}R_{i,j}s_j\right|^2,
\end{align}
that can be computed recursively. Now consider a tree with $M_t$ layers, where the nodes are elements of $\mathcal O$, the branches correspond to the values $d_i$ and each path from the root to a leaf corresponds to a symbol vector $\mathbf s\in\mathcal O^{M_t}$. The ML solution $\mathbf s_{ML}$ is the path with the smallest metric $d_0$. We start from the infinite search radius $r$ and traverse the tree depth-first, updating $r$ whenever a leaf is reached and pruning the branches with $d_i>r$.

The generation of LLRs using \eqref{eq:mlm2} can be decomposed into two steps. The first step is to find the ML solution 
\begin{equation}
\label{eq:ml}
\mathbf s^{ML}=\arg \min_{\mathbf s\in \mathcal O^{M_t}} ||\mathbf{\tilde y}-\mathbf R\mathbf s||^2.
\end{equation}
Assume that the $b$-th bit of its $j$-th symbol is equal to $i$. The second step is to find the best counter-hypothesis
\begin{equation}
\label{eq:counter}
\mathbf s^{\overline{ML}}_{j,b}=\arg \min_{\mathbf s\in \mathcal{X}_{j,b}^{(\overline i)}} ||\mathbf{\tilde y}-\mathbf R\mathbf s||^2,
\end{equation}
i.e. to find the closest to the received signal vector $\mathbf s\in\mathcal O^{M_t}$ that differs from the ML solution in bit $b$ of symbol $s_j$. We use the repeated tree search (RTS) strategy with modified detection order\cite{wang2004approaching,marsh2005smart} and run the tree search $2M\cdot M_t + 1$ times:
\begin{enumerate}
	\item During the first run, we find the ML solution $\mathbf s^{ML}$ and store it.
	\item For each symbol $s_j,j=0,\dots,M_t-1$, we rearrange the columns of $\mathbf H$ so that the detection process starts from layer $j$. Then for each bit $b$ the search starts from the reduced candidate set $\mathcal{X}_{j,b}^{(\overline i)}$ at the first layer and uses the full candidate set $\mathcal O$ for all subsequent layers, which makes the tree pruning efficient and guarantees that only valid counter-hypotheses are considered.
\end{enumerate}
We sort the columns of $\mathbf H$ before each QR decomposition so that the stronger layers are closer to the tree root \cite{wubben2003mmse}.

Another substantial complexity reduction is achieved by LLR clipping \cite{studer2008soft}. Since $\left|L_{j,b}\right|=\left|d(\mathbf s^{ML})-d(\mathbf s^{\overline{ML}}_{j,b})\right|$, after finding $\mathbf s^{ML}$ we can simply initialize the search radius for the subsequent runs as $$r\gets d(\mathbf s^{ML})+\lambda,$$ where $\lambda$ is the clipping value.

\section{Probabilistic shaping for MIMO}
\label{s:mimopas}
\subsection{Receiver design}
Let us start from explaining how to integrate PS into the sphere decoder described in Section \ref{ss:sd}. The amplitude shaping makes paths in the decoding tree non-uniform, which can be accounted for by incorporating a priori probabilities $P(\mathbf s)=\prod_jP(s_j)$ into the search process by adding the $\log P(\mathbf s)$ term to \eqref{eq:ml} and \eqref{eq:counter}. Given that $P(s_j)$ follows MB distribution \eqref{eq:MB}, we obtain 
\begin{align*}
\mathbf s^{ML}&=\arg \min_{\mathbf s\in \mathcal O^{M_t}} \left(||\mathbf{\tilde y}-\mathbf R\mathbf s||^2+\nu ||\mathbf s||^2\right),\\
\mathbf s^{\overline{ML}}_{j,b}&=\arg \min_{\mathbf s\in \mathcal{X}_{j,b}^{(\overline i)}} \left(||\mathbf{\tilde y}-\mathbf R\mathbf s||^2+\nu ||\mathbf s||^2\right),
\end{align*}
and the partial distance computation becomes
$$
d_i(\mathbf s)=d_{i+1}(\mathbf s)+e_i(\mathbf s)+\nu||s_i||^2.
$$
Note that since the underlying constellation for PS is still QAM, the demapper can leverage on its lattice structure to efficiently enumerate the constellation points.

\subsection{Performance}
We evaluate the performance of $4\times4$ MIMO system with 5G LDPC code \footnote{Strictly speaking, 5G LDPC code is not systematic, since a subset of the systematic bits are punctured from the codeword \cite{richardson2018design}. In our evaluations, we apply CCDM only on the set of systematic bits that are transmitted (i.e., not punctured) in order to preserve the shaping.} \cite{3gpp.38.212} in two setups:

\begin{enumerate}
	\item Uniform QAM: take the data bits and encode with $(9600,6400)$ code;
	\item PS: send the data bits to CCDM, obtain $8448$ shaped bits corresponding to the amplitudes that follow MB distribution with $\nu=0.02567$ and then encode with $(9600,8448)$ code.
\end{enumerate}
In both setups, we use layered min-sum belief propagation decoder \cite{hocevar2004reduced} with 25 iterations for decoding LDPC code. Table \ref{tbl:sim} provides the simulation parameters.

\begin{table}[]
	\caption{Simulation assumptions}
	\label{tbl:sim}
	\resizebox{0.9\columnwidth}{!}{%
	\begin{tabular}{|l|l|}
		\hline
		Channel model            & \begin{tabular}[c]{@{}l@{}}TDL-C\cite{3gpp.38.901}\\ 300ns delay spread\\ 11Hz Doppler frequency\end{tabular}            \\ \hline
		Antennas          & 4Tx, 4Rx                                                                                                 \\ \hline
		OFDM symbols          & 1                                                                         \\ \hline
		Resource blocks          & 25                                                                                               \\ \hline
		Information bits  & 6400                                                                                                     \\ \hline
		Block length        & 9600                                                                                                     \\ \hline
		Coding rate         & \begin{tabular}[c]{@{}l@{}}0.6667 for uniform QAM\\ 0.88 for PS (and CCDM with $\nu=0.02567$)\end{tabular}                             \\ \hline
		Modulation          & 256QAM                                                                                                   \\ \hline
		Antenna correlation & \begin{tabular}[c]{@{}l@{}}Low\\ Rx Medium \cite[Sec. 7.7.5.2]{3gpp.38.901} ($\alpha=0,\beta=0.3874$)\end{tabular} \\ \hline
	\end{tabular}
}
\vspace{-0.6cm}
\end{table}

\begin{figure}
	\centering
	\includegraphics[width=0.85\linewidth]{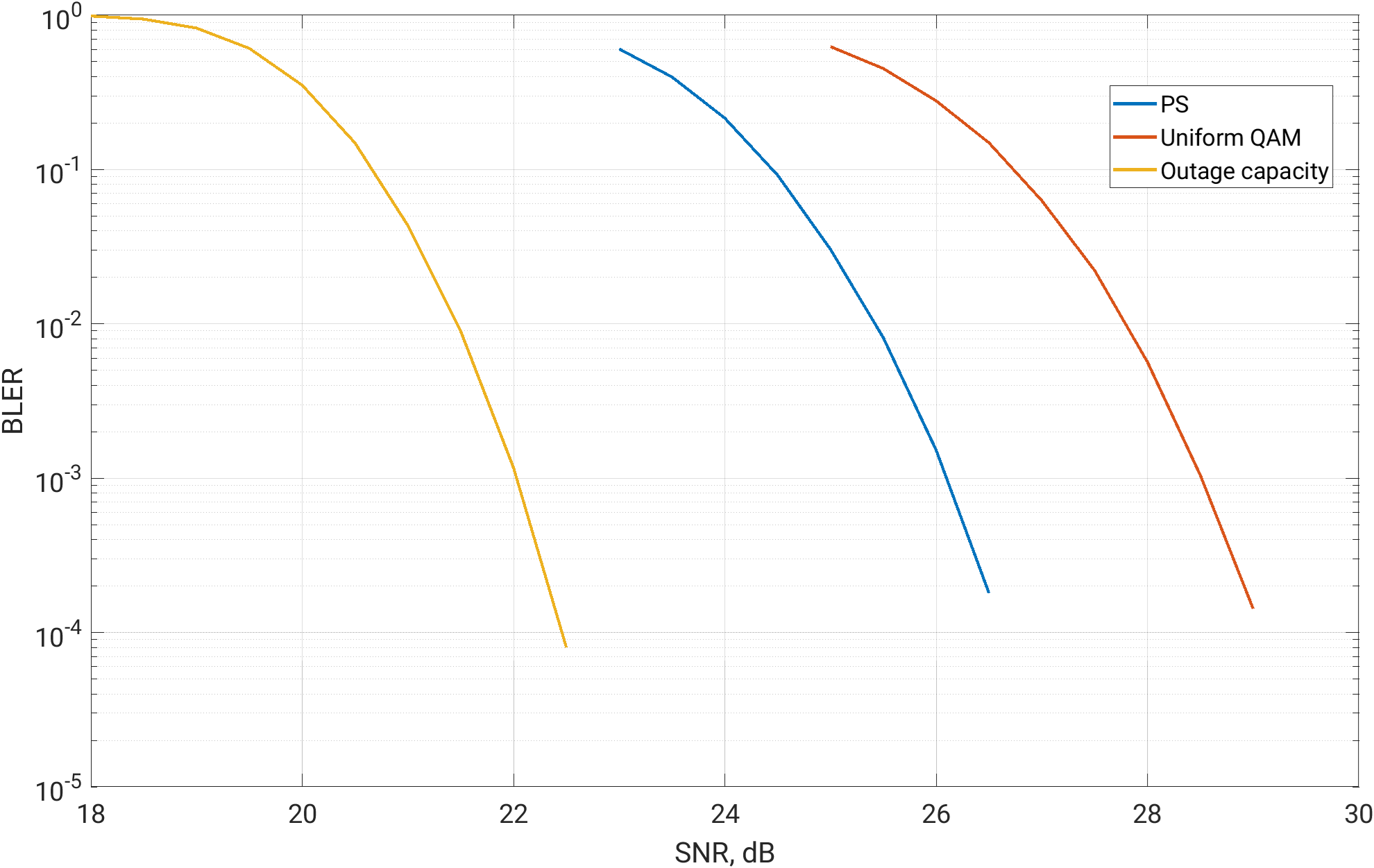}
	\caption{Shaping performance, low correlation}
	\label{fig:pas_perf}
	\vspace{-0.4cm}
\end{figure}

\begin{figure}	
	\centering
	\includegraphics[width=0.85\linewidth]{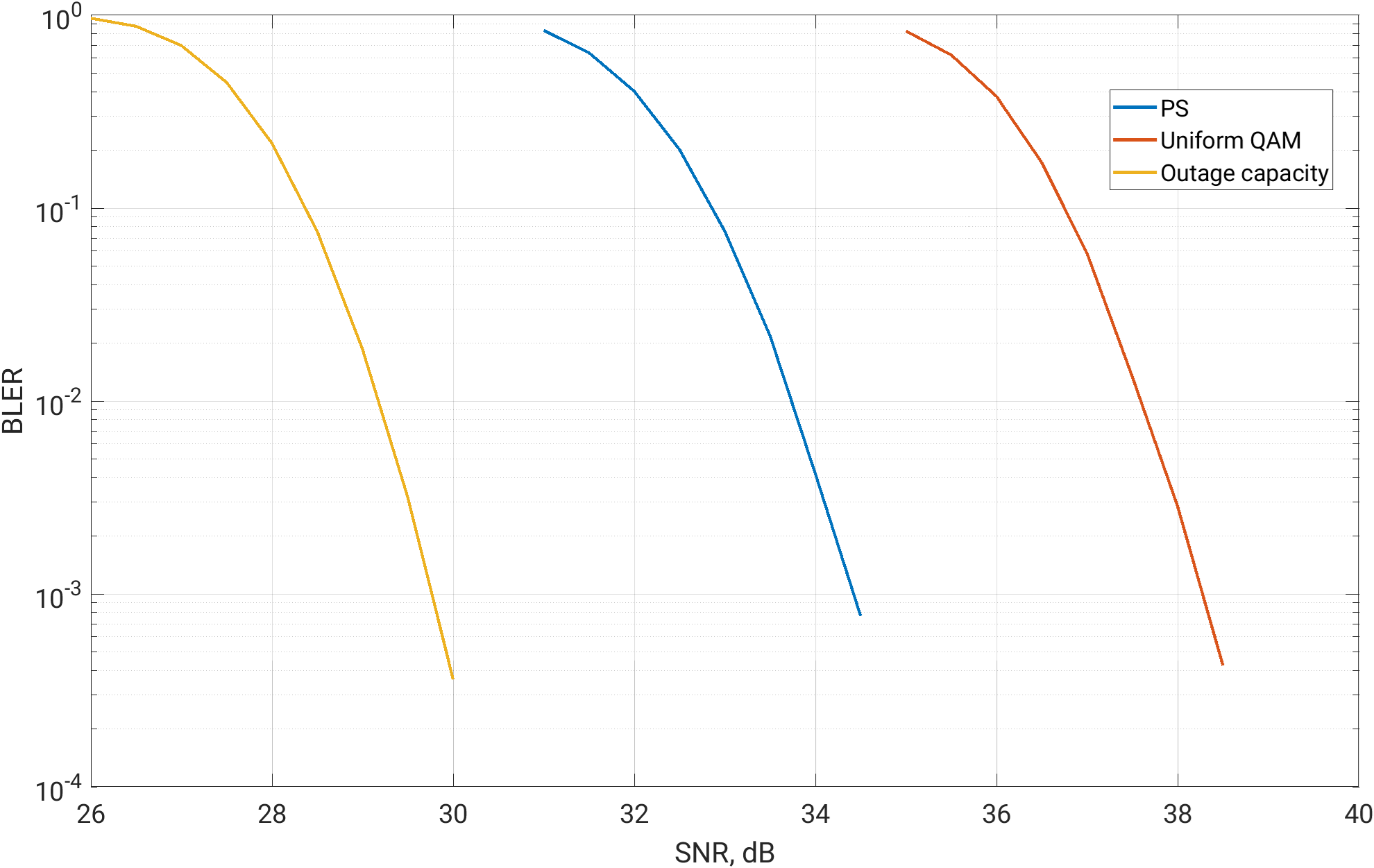}
	\caption{Shaping performance, medium correlation}
	\label{fig:pas_perf_med}
	\vspace{-0.6cm}
\end{figure}

Figure \ref{fig:pas_perf} demonstrates the performance of sphere detector in both setups when the antenna correlation is low. The clipping value $\lambda$ is set to $1000$, which achieves good performance with substantially reduced running time compared to the unbounded search with $\lambda=\infty$. Our results show that probabilistic shaping in MIMO systems can achieve more than 2dB gain, exceeding the 1.53dB AWGN limit. Furthermore, in case of medium antenna correlation the gain from PS gets even larger and reaches 3.5dB, as shown at Figure \ref{fig:pas_perf_med}. This is expected, since in case of medium antenna correlation the inter-layer interference is more detrimental to the performance of MIMO systems and therefore the interference shaping gain from PS gets larger as the antenna correlation increases. To the best of our knowledge, this is the first time when such shaping gains have been reported in the literature. We also plot the outage capacity as a reference. With probabilistic shaping, it becomes possible to operate at 4 dB away from it.

Note that our findings are not limited to the sphere decoder and hold for any (near-)optimal soft-output MIMO demapper as long as a priori probabilities of the constellation symbols are taken into account. 


\subsection{Complexity}
An important aspect regarding the practical implementation of the sphere decoder is its complexity, which is typically determined by the number of nodes visited in the decoding tree. It is a random variable that depends on the received signal and the channel matrix. 

\begin{figure}
	\centering
	\includegraphics[width=0.8\linewidth]{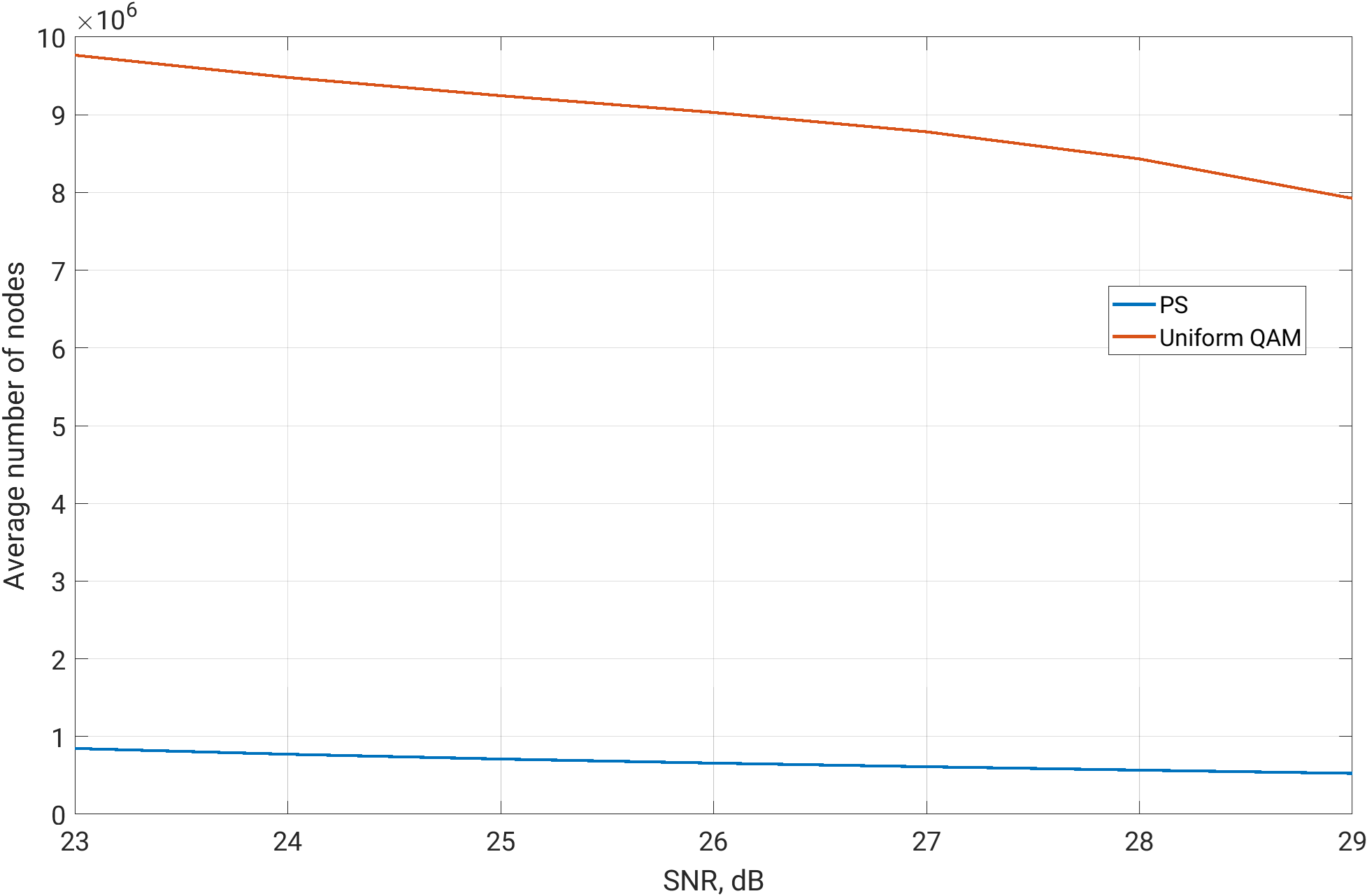}
	\caption{Sphere decoding complexity, low correlation}
	\label{fig:pas_compl}
	\vspace{-0.4cm}
\end{figure}

\begin{figure}
	\centering
	\includegraphics[width=0.8\linewidth]{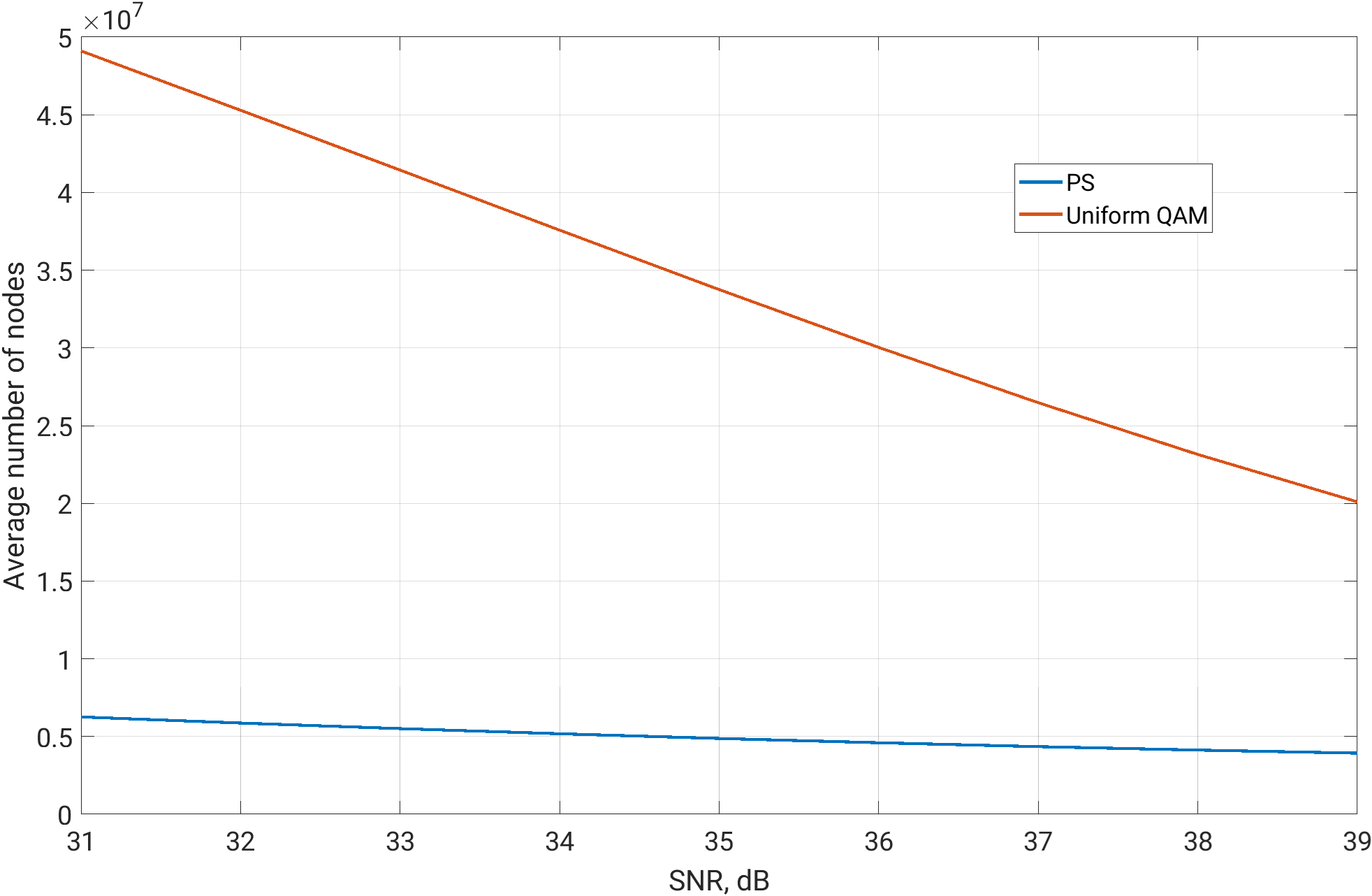}
	\caption{Sphere decoding complexity, medium correlation}
	\label{fig:pas_compl_med}
	\vspace{-0.6cm}
\end{figure}

Figure \ref{fig:pas_compl} demonstrates the average number of nodes visited by the sphere decoder in case of PS and uniform QAM for the same setup as in the previous section. Our results show that utilizing probabilistic shaping reduces the detection complexity by an order of magnitude for the case of low antenna correlation. It appears that incorporating a priori constellation symbol probabilities makes it easier to find the correct path in the decoding tree and exhaust the search. Interestingly, in case of medium antenna correlation the reduction becomes smaller, as shown at Figure \ref{fig:pas_compl_med}. As expected, the medium correlation case is substantially more challenging for sphere decoder from the complexity standpoint, namely the average number of visited nodes grows by a factor of 7x and 3x for the probabilistic shaping and uniform QAM, respectively.

\section{Discussion}
It is important to note that we were able to exceed the AWGN upper bound on the potential shaping gain only with the near-optimal MIMO detector, whereas falling back to linear MMSE method in our evaluations brought the observed performance gains back into the same ballpark as the conventional single-antenna systems with AWGN channel. This result strongly suggests that using a non-linear method is a crucial component but does not give an explanation why.

A promising answer can be found in 3GPP studies of the demodulation performance in multi-user MIMO systems, where the second user, occupying a different subset of spatial layers and treated as interference, can utilize a different modulation order \cite{3gpp.38.878}. It appears that MMSE demodulation performance is insensitive to the QAM order utilized by the second user. On the other hand, ML receiver is able to exploit the interference structure and its performance becomes better when the modulation order of the second user is lower. 

\begin{figure}
	\centering
	\includegraphics[width=0.85\linewidth]{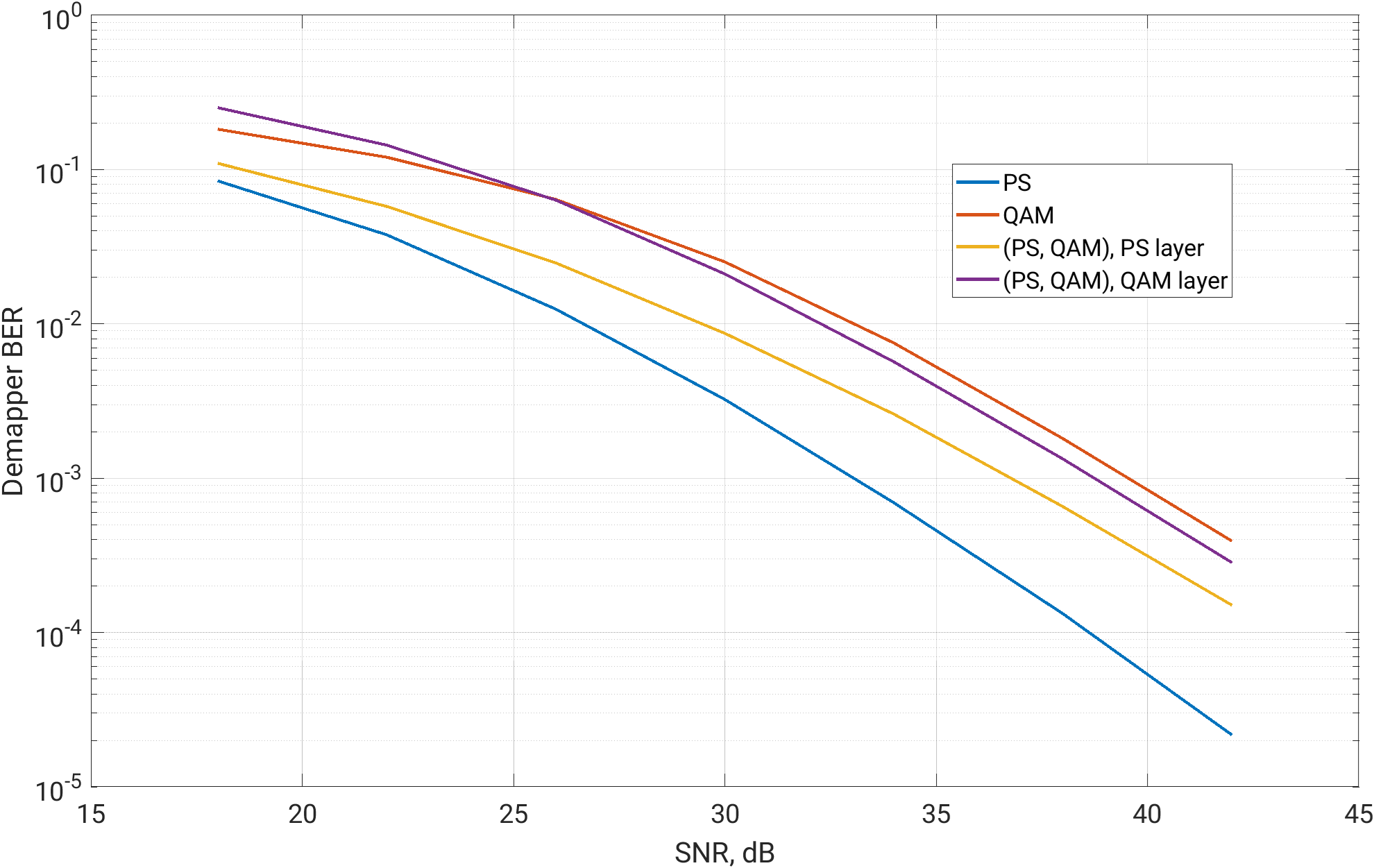}
	\caption{Demapper performance}
	\label{fig:demod_ber}
	\vspace{-0.6cm}
\end{figure}

Let us now look at the sphere decoder output in order to check whether the similar effect holds for the case of probabilistic shaping. We consider $2\times 2$ MIMO system with low antenna correlation for simplicity, and this time we also add the third transmission scenario (PS,QAM), where one spatial layer carries shaped bits and another uniform bits, and we evaluate (BER) for each of the layers. Figure \ref{fig:demod_ber} demonstrates that the shaping gain becomes substantially diminished when the interference layer is unshaped, whereas the demapper performance for unshaped case gets a boost when the interference layer is shaped. This effect becomes more pronounced with a larger number of spatial layers. 

The presented result supports the claim that a good non-linear detector such as sphere decoder makes use of the fact that both the target signal and the interference are shaped, whereas linear MMSE method can only benefit from the power gain due to the non-uniform signaling and as such its achievable gains are limited by the AWGN bound. However, the theoretical limits for probabilistic shaping (and shaping in general) accompanied by the optimal MIMO detector remain unclear and the question whether such scheme is capacity-achievable stays open for future research.

\section{Conclusion}
In this paper, we investigated the application of probabilistic shaping for MIMO channels. Our main finding is that the achievable gains can substantially exceed those theoretically possible for AWGN channel. In the case of low antenna correlation the shaping gain is around 2dB, whereas for the medium correlation scenario it gets even larger, exceeding 3.5dB. In addition, shaping can bring substantial complexity savings for tree-search detection algorithms by reducing the average number of visited nodes.

We would like to emphasize that the crucial component for achieving such gains is to utilize a non-linear demapper, capable of leveraging on the shaped structure of the target signal as well as the interference. Our experiments suggest that reduced-complexity decoders based on the QR decomposition such as sphere decoder are sufficient for that purpose.

We hope that our findings spark the interest in further studying this phenomenon from both practical and theoretical standpoints and pave the road for future investigations into the analysis of shaping in multi-antenna systems.

\enlargethispage{-1.2cm} 

\bibliographystyle{IEEEtran}
\bibliography{biblio}
\end{document}